\documentclass{article}

\usepackage{graphicx}

\def\abstract#1{\vskip 7mm 
        \begin{center}{\large Abstract}\par \smallskip
                \begin{minipage}[c]{12cm}
                        \small #1
                \end{minipage}
        \end{center}
}
\def\title#1{\begin{center}{\Large\bf #1}\end{center}}
\def\author#1{\vskip 5mm \begin{center}{#1}\end{center}}
\def\address#1{\begin{center}{\it #1}\end{center}}
\makeatletter
\@ifundefined{lesssim}{}{}
\@ifundefined{gtrsim}{}{}
\def\vereq#1#2{\lower3pt\vbox{\baselineskip1.5pt \lineskip1.5pt
\ialign{$\m@th#1\hfill##\hfil$\crcr#2\crcr\sim\crcr}}}
\makeatother

\begin{document}

\title{%
  Comparison of Post-Newtonian and Numerical Evolutions\\
  of Black-Hole Binaries
}
\author{%
  Hiroyuki Nakano\footnote{E-mail:hxnsma@rit.edu}, 
  Manuela Campanelli\footnote{E-mail:manuela@astro.rit.edu}, 
  Carlos O. Lousto\footnote{E-mail:colsma@rit.edu},  
  and
  Yosef Zlochower\footnote{E-mail:yrzsma@rit.edu}
}
\address{%
  Center for Computational Relativity and Gravitation, 
  and School of Mathematical Sciences, \\
  Rochester Institute of Technology, Rochester, New York 14623, USA
}

\abstract{
In this paper, we compare the waveforms from 
the post-Newtonian (PN) approach with the 
numerical simulations of generic black-hole binaries 
which have mass ratio $q\sim0.8$, arbitrarily 
oriented spins with magnitudes $S_1/m_1^2\sim0.6$ and $S_2/m_2^2\sim0.4$,
and orbit 9 times from an initial orbital separation of $r\approx11M$ 
prior to merger. 
We observe a reasonably good agreement between the PN and
numerical waveforms, with an overlap of over $98\%$ 
for the first six cycles of 
the $(\ell=2,m=\pm2)$ mode and over $90\%$ for the $(\ell=2,m=1)$ and $(\ell=3,m=3)$ modes. 
}
\section{Introduction}

In 2005, two complementary and independent methods were discovered 
that allowed numerical relativists to completely solve 
the black-hole binary problem in full strong-field 
gravity~\cite{Pretorius:2005gq, Campanelli:2005dd, Baker:2005vv}.
On the other hand, there are currently major experimental 
and theoretical efforts underway to measure these gravitational wave signals.
Therefore, one of the most important tasks of numerical relativity (NR) 
is to assist gravitational wave observatories in detecting gravitational 
waves and extracting the physical parameters of the sources. 
Given the demanding resources required to generate these black-hole 
binary simulations, it is necessary to develop various techniques 
in order to model arbitrary binary configuration based on 
numerical simulations in combination with post-Newtonian (PN) 
and perturbative (e.g. black-hole perturbation) calculations. 

In this paper, we compare the NR and PN waveforms for the challenging problem 
of a generic black-hole binary, i.e., a binary with unequal masses and unequal, 
non-aligned, and precessing spins. 
Comparisons of numerical simulations with post-Newtonian 
ones have several benefits aside from the theoretical verification of PN. 
From a practical point of view, 
one can directly propose a phenomenological description 
and thus make predictions in regions of the parameter space 
still not explored by numerical simulations. 
From the theoretical point of view, an important application is 
to have a calibration of the post-Newtonian error 
in the last stages of the binary merger. 

The paper is organized as follows. In Sec. II we present 
our method to derive the PN gravitational waveforms
from generic black-hole binaries, 
and in III we compare the NR and PN waveforms. 
Finally in Sec. IV we summarize this paper and discuss remaining problems. 
The detailed numerical method and PN calculation presented here 
have been given in \cite{Campanelli:2008nk}. 

\section{Gravitational waveforms in the PN approach}

In order to calculate PN gravitational waveforms, 
we need to calculate the orbital motion of binaries 
in the post-Newtonian approach. 
Here we use the ADM-TT gauge, 
which is the closest to our quasi-isotropic 
numerical initial data coordinates. 
In this paper, we use the PN equations of motion (EOM) based 
on~\cite{Buonanno:2005xu,Damour:2007nc,Steinhoff:2007mb}. 
The Hamiltonian is given in~\cite{Buonanno:2005xu}, 
with the additional terms, i.e., the next-to-leading order gravitational 
spin-orbit and spin-spin couplings 
provided in~\cite{Damour:2007nc,Steinhoff:2007mb}, and
the radiation-reaction force given in~\cite{Buonanno:2005xu}. 
The Hamiltonian which we used here is given by 
\begin{eqnarray}
H &=& H_{\rm O,Newt} + H_{\rm O,1PN} + H_{\rm O,2PN} + H_{\rm O,3PN} 
\nonumber \\ && 
+ H_{\rm SO,1.5PN} + H_{\rm SO,2.5PN} 
+ H_{\rm SS,2PN} + H_{\rm S_1S_2,3PN} \,,
\label{eq:H}
\end{eqnarray}
where the subscript O, SO and SS denote the pure orbital (non-spinning) part, 
spin-orbit coupling and spin-spin coupling, respectively, 
and Newt, 1PN, 1.5PN, etc., refer to the perturbative order in the 
post-Newtonian approach. 
From this Hamiltonian, the conservative part of the orbital and spin EOM 
is derived using the standard techniques of the Hamiltonian formulation. 
For the dissipative part, we use the non-spinning radiation reaction 
results up to 3.5PN (which contributes to the orbital EOM at 6PN order), 
as well as the leading spin-orbit 
and spin-spin coupling to the radiation reaction~\cite{Buonanno:2005xu}.

The above PN evolution is used both to produce very low eccentricity 
orbital parameters at $r\approx11M$ from an initial orbital separation of $50M$, 
and to evolve the orbit from $r\approx11M$. We use these same 
parameters at $r\approx11M$ to generate the initial data for our 
numerical simulations. 
The initial binary configuration at $r=50M$ had 
the mass ratio $q=m_1/m_2 = 0.8$, $\vec S_1/m_1^2 = (-0.2, -0.14,0.32)$, 
and $\vec S_2/m_2^2 =(-0.09, 0.48, 0.35)$.

We then construct a hybrid waveform from the orbital motion 
by using the following procedure. 
First we use the 1PN accurate waveforms derived by 
Wagoner and Will~\cite{Wagoner:1976am} 
(WW waveforms) for a generic orbit. By using these waveforms, 
we can introduce effects due to the black-hole spins, including 
the precession of the orbital plane.  On the other hand, 
Blanchet {\it et al}.~\cite{Blanchet:2008je} 
recently obtained the 3PN waveforms (B waveforms) 
for non-spinning circular orbits. 
We combine these two waveforms to produce a hybrid waveform. 
In order to combine the WW and B waveforms, 
we need to take into account differences in the definitions
of polarization states and the angular coordinates. 
The WW waveforms use the standard definition of GW polarization states, 
which are the same as those derived from the Weyl scalar, 
but the B waveforms use an alternate definition. 
The angular coordinates in the B waveforms are derived from 
circular orbits in the equatorial (xy) plane. To directly 
compare the NR and PN waveforms, we must add a time dependent 
inclination to the B waveforms because in the generic case 
the orbital planes are inclined with respect to the xy plane. 

We note that since there is no gauge ambiguity for combining 
the two waveforms, the combination of the WW and B waveforms 
is unique. Also, it should be noted that we calculate the spin contribution 
to the waveforms through its effect on the orbital motion directly 
in the WW waveforms and indirectly in B waveforms 
through the inclination of the orbital plane. 

For the NR simulations we calculate the Weyl scalar $\psi_4$ 
and then convert the  $(\ell,m)$ modes of $\psi_4$ into $(\ell,m)$ modes of
$h = h_{+} - i h_{\times}$. 

\section{Comparison of the NR and PN waveforms}

To compare PN and numerical waveforms, we need to determine 
the time translation $\delta t$ between the numerical time and 
the corresponding point on the PN trajectory. That is to say, 
the time it takes for the signal to reach the extraction sphere 
($r=100M$ in our numerical simulation). 
We determine this by finding the time translation near $\delta t=100M$ 
that maximizes the agreement of the early time waveforms 
in the $(\ell=2,m=\pm2)$, $(\ell=2,m=\pm1)$, and 
$(\ell=3,m=\pm3)$ simultaneously. We find $\delta t \sim 112$, in
good agreement with the expectation for our observer at $r=100M$. 
Since our PN waveforms are given uniquely 
by a binary configuration, i.e., an actual location of the PN particle, 
we do not have any time shift or phase modification 
other than this retardation of the signal. 
It is noted that other methods which are not based on the particle locations, 
have freedom in choosing a phase factor. 

\begin{figure}[ht]
\center
\includegraphics[width=2.9in]{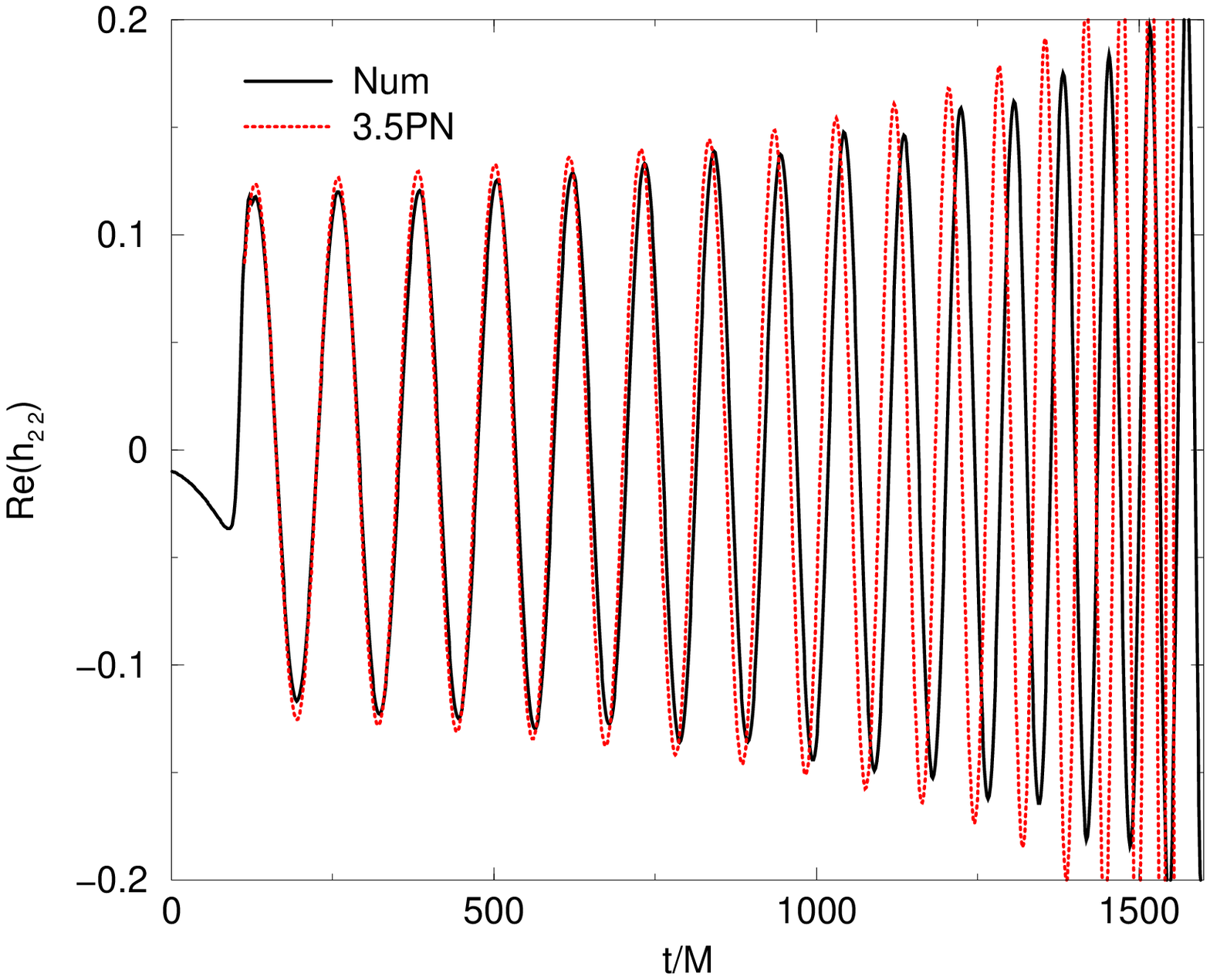}
\includegraphics[width=2.9in]{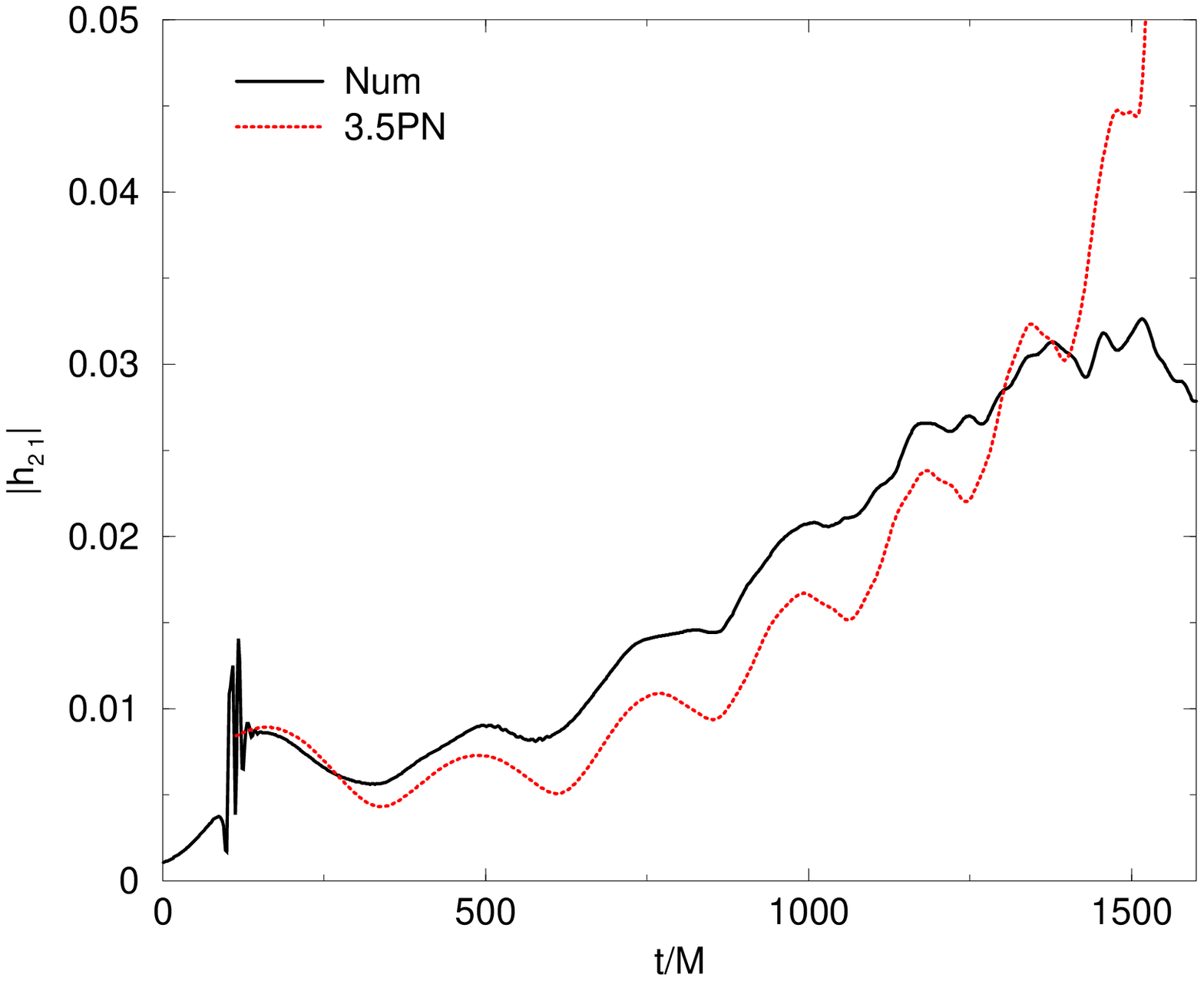}
\caption{
{\bf Left:} The real part of the $(\ell=2, m=2)$ mode of $h$ 
from the numerical and 3.5PN simulations. 
Here 3.5PN predicts an early merger and has a higher frequency 
than the numerical waveform. 
{\bf Right:} The amplitude of the $(\ell=2, m=1)$ mode of $h$ 
from the numerical and 3.5PN simulations.
The secular oscillation in the numerical amplitude occurs 
at roughly the precessional frequency. The timescale is of order $1000M$. 
Here the shorter-timescale oscillations correspond roughly to the orbital 
period. 
}
\label{fig:G3.5}
\end{figure}

In the left panel of Fig.~\ref{fig:G3.5}, 
we show the real part of the $(\ell=2, m=2)$ 
mode of the strain $h$ with this time translation. 
(The other modes are shown in~\cite{Campanelli:2008nk}.)
We note that the reasonable agreement of the numerical 
and PN waveforms for $700M$. 

From the analysis of the amplitudes of each mode, 
we see that the precession and eccentricity of the orbit 
impart signatures 
on the modes of the waveform at the orbital frequency. 
However, the long-time oscillations in the amplitudes, 
here apparent only in the ($\ell = 2,\,m = \pm 1$) modes, 
seem to be due purely to precession, and occur 
at the precessional frequency. 
In the right panel of Fig.~\ref{fig:G3.5}, 
we show the amplitudes of the $(\ell=2,m=1)$ mode of $h$. 

Next, in order to quantitatively compare the modes of the PN waveforms 
with the numerical waveforms we define the overlap, 
or matching criterion, for the real and imaginary parts of
each mode as 
\begin{eqnarray}
  \label{eq:match}
  M_{\ell m}^\Re = \frac{<R^{\rm Num}_{\ell m},
    R^{\rm PN}_{\ell m}>}
     {\sqrt{<R^{\rm Num}_{\ell m},R^{\rm Num}_{\ell m}>
     <R^{\rm PN}_{\ell m},R^{\rm PN}_{\ell m}>}} \,, \quad 
  M_{\ell m}^\Im = \frac{<I^{\rm Num}_{\ell m},
    I^{\rm PN}_{\ell m}>}{\sqrt{<I^{\rm Num}_{\ell m},
    I^{\rm Num}_{\ell m}><I^{\rm PN}_{\ell m},I^{\rm PN}_{\ell m}>}} \,,
\end{eqnarray}
where
$R_{\ell m}$ and $I_{\ell m}$ are defined 
by the real and imaginary parts of the waveform mode $h_{\ell m}$, 
respectively, 
and the inner product is calculated by 
$ 
<f,g> = \int_{t_1}^{t_2} f(t) g(t) dt
$. 
Hence, $M_{\ell m}^\Re = M_{\ell m}^\Im = 1$ indicates that the given
PN and numerical mode agree. The results of 
these matching studies are summarized in Table~\ref{tab:G3.5match}. 

\begin{table}[ht]
  \caption{The match of the real and imaginary parts of the modes of $h$
of the G3.5 configuration for the 3.5 PN waveforms 
and the numerical waveforms with the time translation $\delta t = 112.5$.}
  \label{tab:G3.5match}
 \center
\renewcommand{\arraystretch}{1.2}
  \begin{tabular}{l||c|c|c}
   Integration Time & 600 & 800 & 1000 \\
\hline
  $M_{22}^\Re$ & 0.986 & 0.964 & 0.895 \\
  $M_{22}^\Im$ & 0.987 & 0.962 & 0.900 \\
  $M_{2-2}^\Re$ & 0.986 & 0.964 & 0.895 \\
  $M_{2-2}^\Im$ & 0.987 & 0.962 & 0.901 \\

  $M_{21}^\Re$ & 0.904 & 0.912 & 0.843 \\
  $M_{21}^\Im$ & 0.916 & 0.901 & 0.820 \\
  $M_{2-1}^\Re$ & 0.920 & 0.908 & 0.833 \\
  $M_{2-1}^\Im$ & 0.917 & 0.903 & 0.816 \\

  $M_{33}^\Re$ & 0.938 & 0.891 & 0.738 \\
  $M_{33}^\Im$ & 0.919 & 0.868 & 0.721 \\
  $M_{3-3}^\Re$ & 0.931 & 0.880 & 0.733 \\
  $M_{3-3}^\Im$ & 0.906 & 0.857 & 0.721 \\
  \end{tabular}
\end{table}

We also determine an alternate time translation, one wavelength in
the $(\ell=2,m=2)$ mode, that increases the matching of 
the $(\ell=2,m=2)$ mode over longer integration periods. 
On the other hand, this new time translation, $\delta t = 233$, 
causes the $(\ell=3)$ modes to be out of phase, 
leading to negative overlaps. Thus by looking at the
$(\ell=2)$ and $(\ell=3)$ modes simultaneously, 
we can reject this false match. 

\section{Conclusion and discussion}

We analyzed the first long-term generic waveform produced 
by the merger of unequal mass, unequal spins, precessing black holes. 
It is found that a good initial agreement of waveforms 
for the first six cycles, with overlaps of over $98\%$ 
for the $(\ell=2, m=\pm2)$ modes, over $90\%$ 
for the $(\ell=2, m=\pm1)$ modes, and over $90\%$ 
for the $(\ell=3, m=\pm3)$ modes. 
These agreement degrades as 
we approach the more dynamical region of the late merger and plunge.

There are some remaining problems. 
The PN gravitational waveforms used here do not include direct spin effects. 
We considered the spin contribution to the waveform 
only through its effect on the orbital motion. Recently, 
the direct spin effects have been discussed in~\cite{Arun:2008kb}.  
And also in the PN approach, 
the waveforms are derived from binaries whose each body is 
considered as a point particle. The finite size effects 
of the bodies is also important in the late-inspiral region. 
Furthermore, we will need higher-order post-Newtonian 
calculations of both spin-orbit and spin-spin terms, 
especially for the phase evolution of gravitational waves. 

We also have a important issue. 
In order to detect the gravitational waves from binaries, 
it is necessary to study the data analysis. 
(For example, the Numerical INJection Analysis (NINJA) project~\cite{ninja}.) 
Here, we must treat a very large parameter space for intrinsic parameters 
of black-hole binaries, 
The development of effective GW templates for the whole history 
of binaries, i.e., the inspiral, merger and ringdown should be done.

\subsection*{Acknowledgments}

We would like to thank H.~Tagoshi and R.~Fujita for useful discussions.


\begin{thebibliography}{99}

\bibitem{Pretorius:2005gq}
  F.~Pretorius,
  Phys.\ Rev.\ Lett.\  {\bf 95}, 121101 (2005)
  [arXiv:gr-qc/0507014].

\bibitem{Campanelli:2005dd}
  M.~Campanelli, C.~O.~Lousto, P.~Marronetti and Y.~Zlochower,
  Phys.\ Rev.\ Lett.\  {\bf 96}, 111101 (2006)
  [arXiv:gr-qc/0511048].

\bibitem{Baker:2005vv}
  J.~G.~Baker, J.~Centrella, D.~I.~Choi, M.~Koppitz and J.~van Meter,
  Phys.\ Rev.\ Lett.\  {\bf 96}, 111102 (2006)
  [arXiv:gr-qc/0511103].

\bibitem{Campanelli:2008nk}
  M.~Campanelli, C.~O.~Lousto, H.~Nakano and Y.~Zlochower,
  arXiv:0808.0713 [gr-qc].

\bibitem{Buonanno:2005xu}
  A.~Buonanno, Y.~Chen and T.~Damour,
  Phys.\ Rev.\  D {\bf 74}, 104005 (2006)
  [arXiv:gr-qc/0508067].

\bibitem{Damour:2007nc}
  T.~Damour, P.~Jaranowski and G.~Schaefer,
  Phys.\ Rev.\  D {\bf 77}, 064032 (2008)
  [arXiv:0711.1048 [gr-qc]].

\bibitem{Steinhoff:2007mb}
  J.~Steinhoff, S.~Hergt and G.~Schaefer,
  Phys.\ Rev.\  D {\bf 77}, 081501 (2008)
  [arXiv:0712.1716 [gr-qc]].

\bibitem{Wagoner:1976am}
  R.~V.~Wagoner and C.~M.~Will,
  Astrophys.\ J.\  {\bf 210} (1976) 764
  [Erratum-ibid.\  {\bf 215} (1977) 984].

\bibitem{Blanchet:2008je}
  L.~Blanchet, G.~Faye, B.~R.~Iyer and S.~Sinha,
  arXiv:0802.1249 [gr-qc].

\bibitem{Arun:2008kb}
  K.~G.~Arun, A.~Buonanno, G.~Faye and E.~Ochsner,
  arXiv:0810.5336 [gr-qc].

\bibitem{ninja}
Numerical INJection Analysis (NINJA) Project Home Page, \\
http://www.gravity.phy.syr.edu/dokuwiki/doku.php?id=ninja:home



\end{thebibliography}
\end{document}